\documentclass[fleqn,usenatbib]{mnras}

\usepackage{newtxtext,newtxmath}

\usepackage[T1]{fontenc}

\DeclareRobustCommand{\VAN}[3]{#2}
\let\VANthebibliography\thebibliography
\def\thebibliography{\DeclareRobustCommand{\VAN}[3]{##3}\VANthebibliography}

\usepackage{graphicx}	
\usepackage{amsmath}	

\newcommand{\Msun}{\ensuremath{\mathrm{M}_\odot}}

\newcommand{\Lsun}{\ensuremath{\mathrm{L}_\odot}}

\newcommand{\Mej}{\ensuremath{M_\mathrm{ej}}}
\newcommand{\Eej}{\ensuremath{E_\mathrm{ej}}}
\newcommand{\Msunpyr}{\ensuremath{\Msun~\mathrm{yr^{-1}}}}
\newcommand{\kmps}{\ensuremath{\mathrm{km~s^{-1}}}}

\graphicspath{{./}{figures/}}

\title[RSG mass loss and SN radio properties]{
Constraining red supergiant mass-loss prescriptions through supernova radio properties
}

\author[T. J. Moriya]{
Takashi J. Moriya$^{1,2}$\thanks{E-mail: takashi.moriya@nao.ac.jp (TJM)}
\\
$^{1}$National Astronomical Observatory of Japan, National Institutes of Natural Sciences, 2-21-1 Osawa, Mitaka, Tokyo 181-8588, Japan \\
$^{2}$School of Physics and Astronomy, Faculty of Science, Monash University, Clayton, Victoria 3800, Australia
}

\date{Accepted 2021 February 13. Received 2021 February 13; in original form 2021 January 13}

\pubyear{2021}

\begin{document}
\label{firstpage}
\pagerange{\pageref{firstpage}--\pageref{lastpage}}
\maketitle

\begin{abstract}
Supernova properties in radio strongly depend on their circumstellar environment and they are an important probe to investigate the mass loss of supernova progenitors. Recently, core-collapse supernova observations in radio have been assembled and the rise time and peak luminosity distribution of core-collapse supernovae at 8.4 GHz has been estimated. In this paper, we constrain the mass-loss prescriptions for red supergiants by using the rise time and peak luminosity distribution of Type II supernovae in radio. We take the de Jager and van Loon mass-loss rates for red supergiants, calculate the rise time and peak luminosity distribution based on them, and compare the results with the observed distribution. We found that the de Jager mass-loss rate explains the widely spread radio rise time and peak luminosity distribution of Type II supernovae well, while the van Loon mass-loss rate predicts a relatively narrow range for the rise time and peak luminosity. We conclude that the mass-loss prescriptions of red supergiants should have strong dependence on the luminosity as in the de Jager mass-loss rate to reproduce the widely spread distribution of the rise time and peak luminosity in radio observed in Type II supernovae.
\end{abstract}

\begin{keywords}
supernovae: general -- radio continuum: transients -- circumstellar matter -- stars: massive -- stars: mass-loss -- supergiants
\end{keywords}



\section{Introduction}
Red supergiants (RSGs) are progenitors of Type~II supernovae (SNe, \citealt{smartt2015} for a review) that are the most commonly observed core-collapse SNe \citep[][]{shivvers2017frac}. While the upper mass limit of RSGs is found to be around 25~\Msun, the maximum mass of the RSG SN progenitors estimated based on the progenitors observationally identified so far is around $17~\Msun$ \citep[e.g.,][]{kochanek2008,smartt2009}. The possible reasons for the discrepancy in the maximum masses of RSGs and RSG SN progenitors have been actively discussed \citep[e.g.,][]{horiuchi2011,walmswell2012rsgdust,adams2017,sukhbold2020}, although the exact maximum mass of the RSG SN progenitors could be higher than 17~\Msun\ and the maximum mass difference may not be statistically significant yet \citep[][]{davies2020}. In particular, the uncertainties in RSG mass loss have been suggested to be a possible reason for the discrepancy \citep[e.g.,][]{yoon2010,georgy2012}. Several empirical mass-loss prescriptions have been proposed for RSGs (e.g., \citealt{mauron2011} for a summary) and they have a strong influence on the predicted properties of SN progenitors \citep[e.g.,][]{meynet2015}.

Matter lost from SN progenitors exists around the progenitors and forms circumstellar matter (CSM). Thus, SN explosions occur within the CSM that imprints the mass-loss history of the SN progenitors. Especially, radio emission from SNe, which is dominated by synchrotron emission from the forward shock, is strongly affected by the CSM properties \citep[e.g.,][]{chevalier1982radiox,chevalier1998ssa} and it has long been used to probe the CSM properties of the SN progenitors \citep[e.g.,][]{weiler2002}. In other words, radio observations from SNe~II can be used to constrain the mass-loss properties of RSGs \citep[][]{chevailer2006}. In addition to radio observations, early SN observations in optical and ultraviolet are starting to unveil the existence of the dense CSM at the immediate vicinity of RSG SN progenitors \citep[e.g.,][]{yaron2017,forster2018}. The dense CSM could be linked to unknown RSG mass-loss mechanism connected to core collapse \citep[e.g.,][]{quataert2012,fuller2017}. In this paper, however, we focus on the distant CSM formed through the canonical RSG mass loss probed by the late SN radio observations.

Although SN radio light curves (LCs) are known to have some modulations \citep[e.g.,][]{weiler1992,soderberg2006}, they are generally characterized by a single-peak LC. The time of the radio LC peak corresponds to the time when the synchrotron self-absorption or the free-free absorption optical depth becomes around unity \citep[][]{chevalier1998ssa}. Therefore, the rise time ($t_\mathrm{rise}$) and peak luminosity ($L_\mathrm{peak}$) of the SN radio LCs depend strongly on the CSM density. \citet{bietenholz2020} recently compiled a large sample of core-collapse SN observations at $2-10~\mathrm{GHz}$ and estimated the rise time and peak luminosity distributions of SNe at around 8.4~GHz. They found that the rise time and peak luminosity distribution of SNe~II (excluding Type~IIn SNe) is characterized by lognormal distributions. Excluding SN~1987A, which is the explosion of a blue supergiant, they found $\log (t_\mathrm{rise}/\mathrm{days}) = 1.7\pm 1.0$ and $\log (L_\mathrm{peak}/\mathrm{erg~s^{-1}~Hz^{-1}}) = 25.4\pm 1.2$ for SNe~II.

In this paper, we aim to constrain the RSG mass-loss prescriptions by using the radio rise time and peak luminosity relation of SNe~II estimated by \citet{bietenholz2020}. The rest of the paper is organized as follows. We first formulate the radio emission from SNe in Section~\ref{sec:radiolc}. We then indroduce the RSG mass-loss prescriptions we investigate in this paper in Section~\ref{sec:rsgmasslosspre}. We compare the rise time and peak luminosity distributions expected from these RSG mass-loss prescriptions with that constrained from the observations and discuss implications in Section~\ref{sec:comparison}. We summarize this paper in Section~\ref{sec:summary}.

\begin{figure}
\includegraphics[width=\columnwidth]{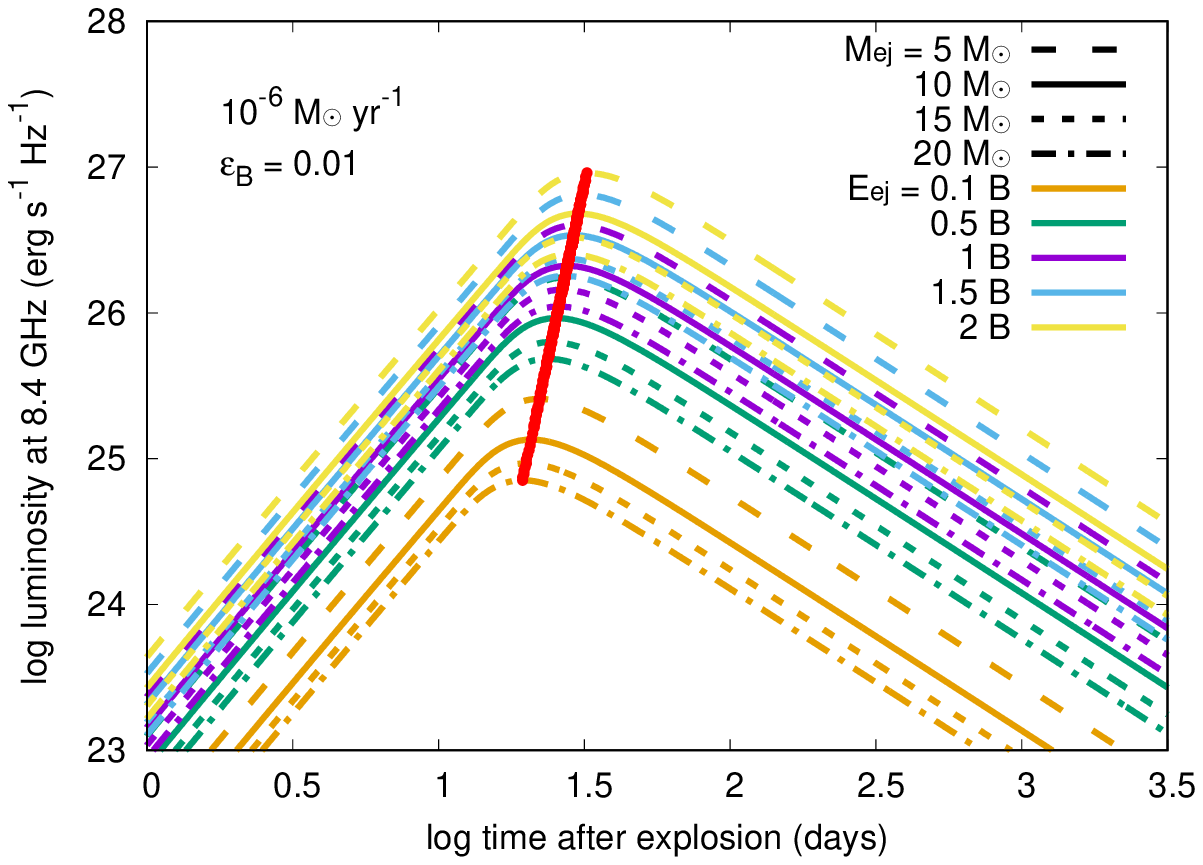}
\caption{
Radio luminosity evolution from SNe in the $\dot{M}=10^{-6}~\Msunpyr$ CSM with $\varepsilon_B = 0.01$ and SSA. The radio LCs with $\Mej=5,10,15$, and 20~\Msun\ and $\Eej=0.1,0.5,1,1.5,$ and 2~B are presented. The red line presents the locations of the LC peaks from all the combinations of $\Mej=5-20~\Msun$ and $\Eej=0.1-2~\mathrm{B}$.
}
\label{fig:m6example}
\end{figure}

\section{Radio luminosity evolution}\label{sec:radiolc}
We first formulate the SN radio luminosity evolution. We assume that synchrotron emission from electrons accelerated at the forward shock by the amplified magnetic field is the dominant radio luminosity source in SNe. The synchrotron emission from the forward shock can be formulated as
\begin{equation}
\nu L_\nu \sim \pi r_\mathrm{sh}^2 v_\mathrm{sh} n_\mathrm{rel}
\left(\frac{\gamma_\nu}{\gamma_\mathrm{min}}\right)^{1-p}
\gamma_\nu m_e c^2
\left[
1+\frac{t_{\mathrm{sync},\nu}}{t}
\right]^{-1},
\label{eq:radiolum}
\end{equation}
where $r_\mathrm{sh}$ is the forward shock radius, $v_\mathrm{sh}$ is the forward shock velocity, $n_\mathrm{rel}$ is the number density of accelerated relativistic electrons, $\gamma_\nu=(2\pi m_ec\nu/eB)^{0.5}$ is the Lorentz factor of the relativistic electrons with the characteristic frequency of $\nu$, $\gamma_\mathrm{min}$ is the minimum Lorentz factor of the accelerated electrons, $m_e$ is the electron mass, $t_\mathrm{sync,\nu}=6\pi m_e c/\sigma_T\gamma_\nu B^2$ is the synchrotron cooling timescale, $e$ is the electron charge, $B$ is the magnetic field strength, and $\sigma_T$ is the Thomson scattering cross section. We assume $\gamma_\mathrm{min}\sim 1$ and $n_\mathrm{rel}\propto \gamma^{-p}$ with $p=3$ \citep[e.g.,][]{chevalier2006fransson,maeda2013}. Because we use the radio LC properties at 8.4~GHz compiled by \citet{bietenholz2020}, all the radio luminosity evolution we calculate in this paper is at 8.4~GHz.

Major uncertain parameters in estimating the radio luminosity are the fraction of the shock kinetic energy used to the electron acceleration ($\varepsilon_e$) and the fraction converted to magnetic energy ($\varepsilon_B$). We adopt $\varepsilon_e = 0.1$ and $\varepsilon_B = 0.01- 0.001$ in this work following the constraints on these parameters based on multi-frequency SN observations \citep[e.g.,][]{bjornsson2004,maeda2012sn2011dhrx,kamble2016sn2013dfrx}.

The synchrotron emission (Eq.~\ref{eq:radiolum}) is affected by absorption processes. We first assume that the synchrotron self-absorption (SSA) is the dominant absorption process \citep{chevalier1998ssa}. The SSA optical depth is $\tau_\mathrm{SSA}=(\nu/\nu_\mathrm{SSA})^{-(p+4)/2}$, where $\nu_\mathrm{SSA}\simeq 3\times 10^{5}(r_\mathrm{sh}\epsilon_e/\epsilon_B)^{2/7}B^{9/7}$~Hz in the cgs unit for $p=3$. 
The free-free absorption can also dominate the absorption process especially when the CSM density is high \citep[e.g.,][]{chevalier1982radiox,chevalier1998ssa,chevailer2006,weiler2002,bietenholz2020}. We discuss the effect of the free-free absorption at the end of Section~\ref{sec:comparison}.

The hydrodynamic properties of the forward shock ($r_\mathrm{sh}$ and $v_\mathrm{sh}$) are estimated based on the formulation in \citet{chevalier1982selfsimilar}. The SN ejecta are assumed to have the two-component power-law density structure ($\propto r^{-n}$ outside and $\propto r^{-\delta}$ inside) with $n=12$ and $\delta =1$ \citep[][]{matzner1999}. The SN explosion energy is varied from 0.1~B to 2~B, where $1~\mathrm{B} \equiv 10^{51}~\mathrm{erg}$, and the SN ejecta mass is set from 5~\Msun\ to 20~\Msun\ \citep[e.g.,][]{pejcha2015}.

Our interest in this paper is in the canonical RSG mass loss before the onset of the possible mass-loss enhancement shortly before RSG SN explosions. Thus, we assume that the CSM structure is formed through steady mass loss and has the following density structure
\begin{equation}
    \rho_\mathrm{CSM}(r) = \frac{\dot{M}}{4\pi v_\mathrm{CSM}}r^{-2},
\end{equation}
where $\dot{M}$ is the mass-loss rate and $v_\mathrm{CSM}$ is the wind velocity. We set $v_\mathrm{CSM}=20~\kmps$, which is typically observed in RSGs \citep[][]{goldman2017}. 

Combining the assumptions presented so far, we can estimate the radio luminosity evolution for a given $\dot{M}$. Fig.~\ref{fig:m6example} shows the radio luminosity evolution for $\dot{M}=10^{-6}~\Msunpyr$ with some combinations of \Mej\ and \Eej\ adopting only SSA.
For a given $\dot{M}$, the rise time and peak luminosity make a single line when we plot the radio LCs with all the combinations of $\Mej = 5-20~\Msun$ and $\Eej = 0.1-2~\mathrm{B}$. An example is shown in Fig.~\ref{fig:m6example} as the red line in the case of $\dot{M}=10^{-6}~\Msunpyr$.

\begin{figure}
\includegraphics[width=\columnwidth]{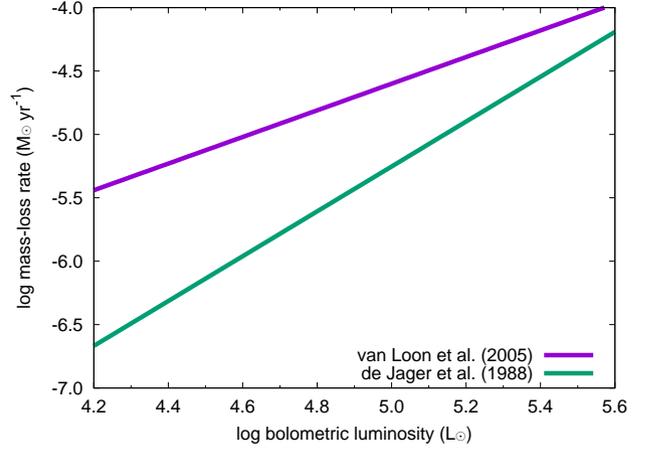}
\caption{
Empirical mass-loss rates of RSGs as a function of their bolometric luminosity. The effective temperature of 3500~K is assumed. We adopt the two representative mass-loss prescriptions of \citet{vanloon2005} and \citet{dejager1988}.
}
\label{fig:masslossrate}
\end{figure}

\begin{figure}
\includegraphics[width=\columnwidth]{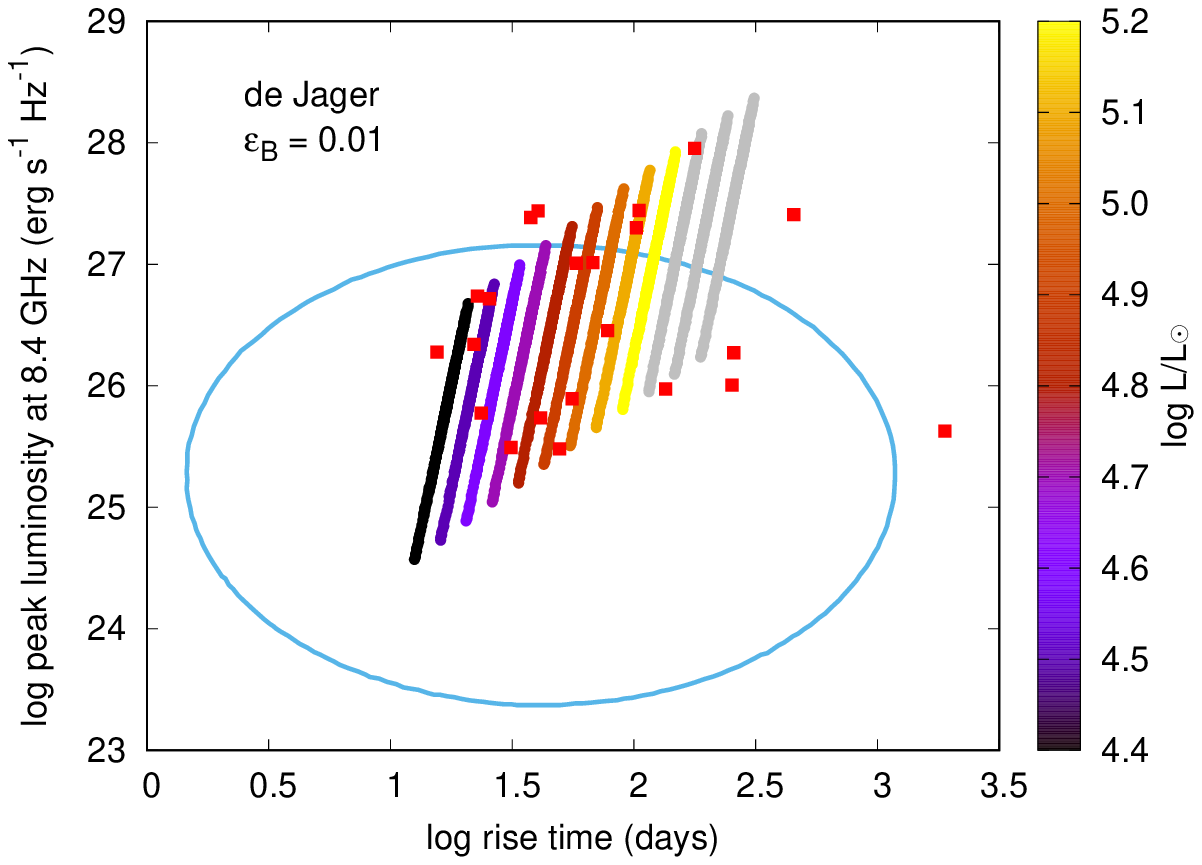}
\includegraphics[width=\columnwidth]{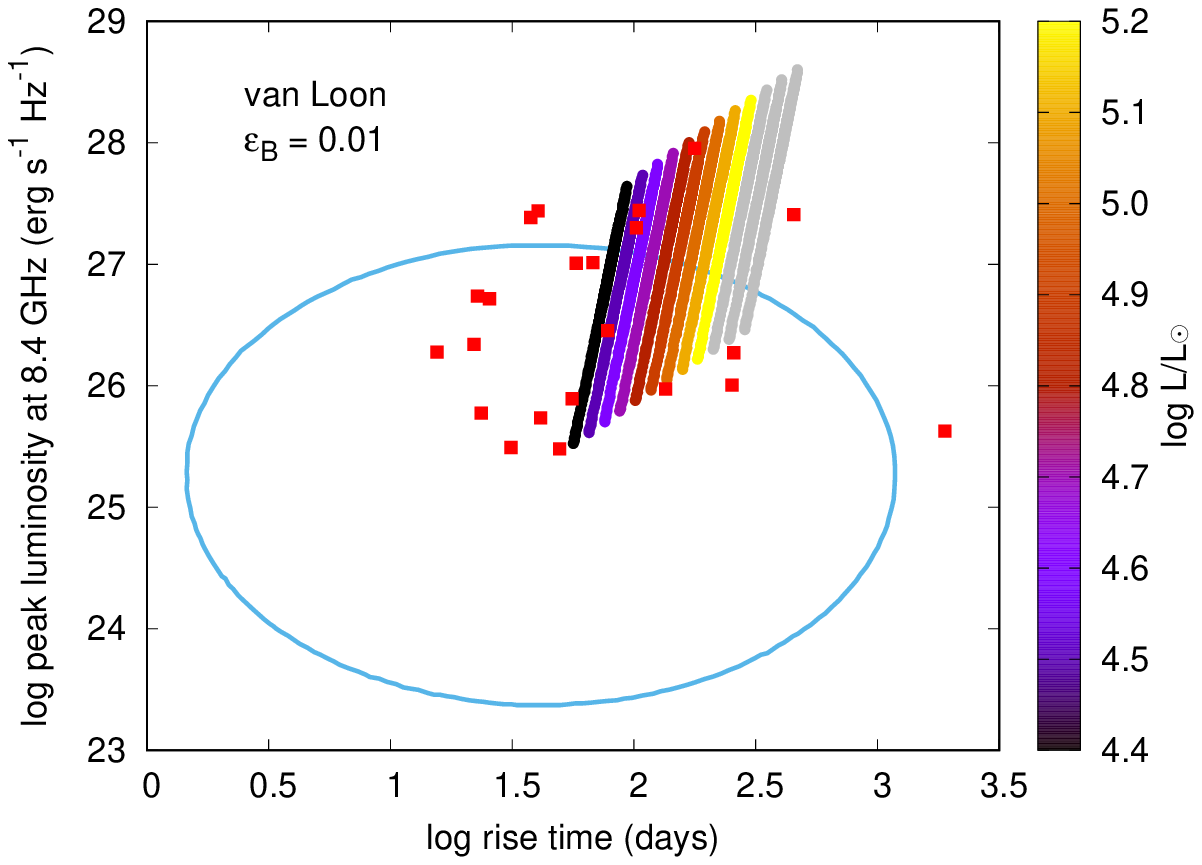}
\caption{
Rise time and peak luminosity of SNe~II expected from the de Jager mass-loss rate (top) and the van Loon mass-loss rate (bottom). $\varepsilon_B=0.01$ and SSA are adopted. The colored lines show those for the RSG SN progenitor luminosity range ($\log L/\Lsun = 4.4-5.2$). For a given luminosity, each line shows the rise time and peak luminosity range for $\Mej=5-20~\Msun$ and $\Eej=0.1-2~\mathrm{B}$ (cf. Fig.~\ref{fig:m6example}). The gray lines show RSGs with $\log L/\Lsun = 5.3, 5.4,$ and 5.5 which may not explode as SNe~II. The red squares show the rise time and peak luminosity of SNe~II obtained in \citet[][]{bietenholz2020}. The blue ellipse shows the region where the 68\% of SNe~II are estimated to exist based on all the observational information available including upper limits \citep[][]{bietenholz2020}.
}
\label{fig:standard}
\end{figure}

\begin{figure}
\includegraphics[width=\columnwidth]{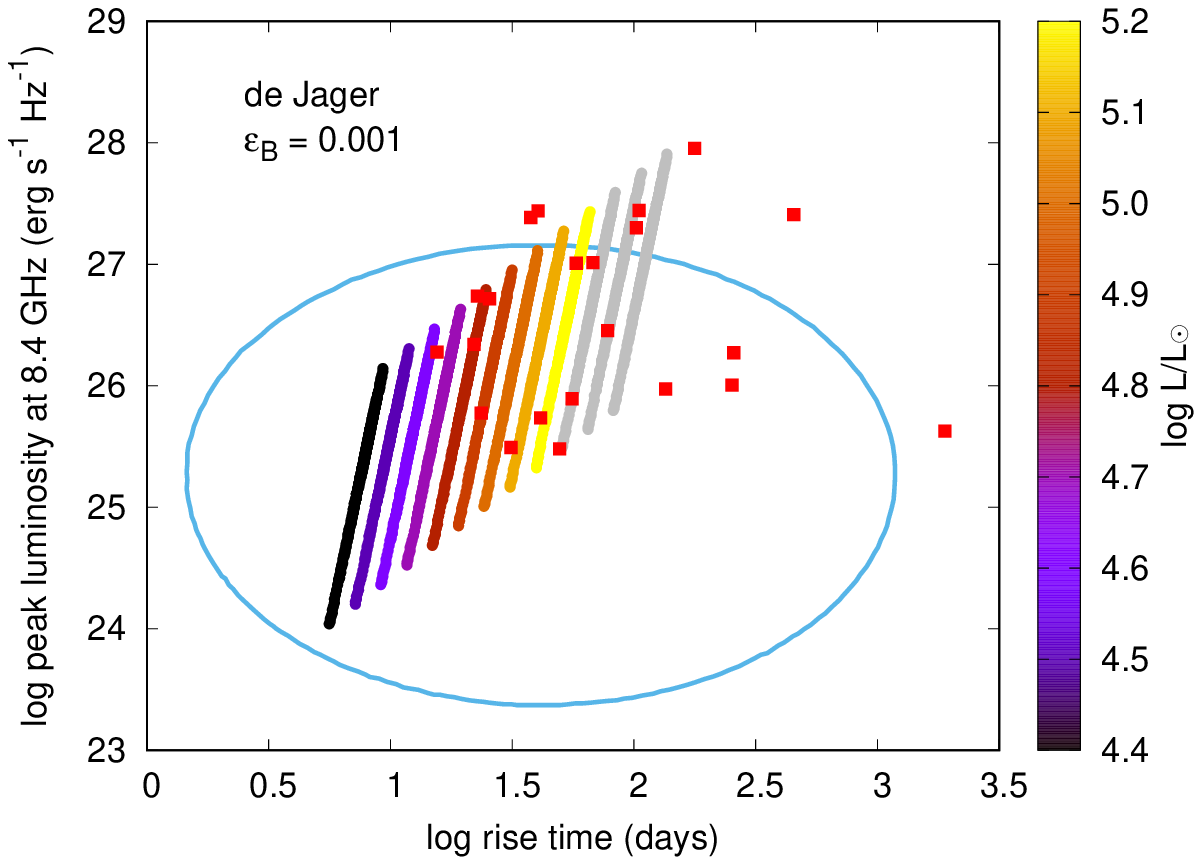}
\includegraphics[width=\columnwidth]{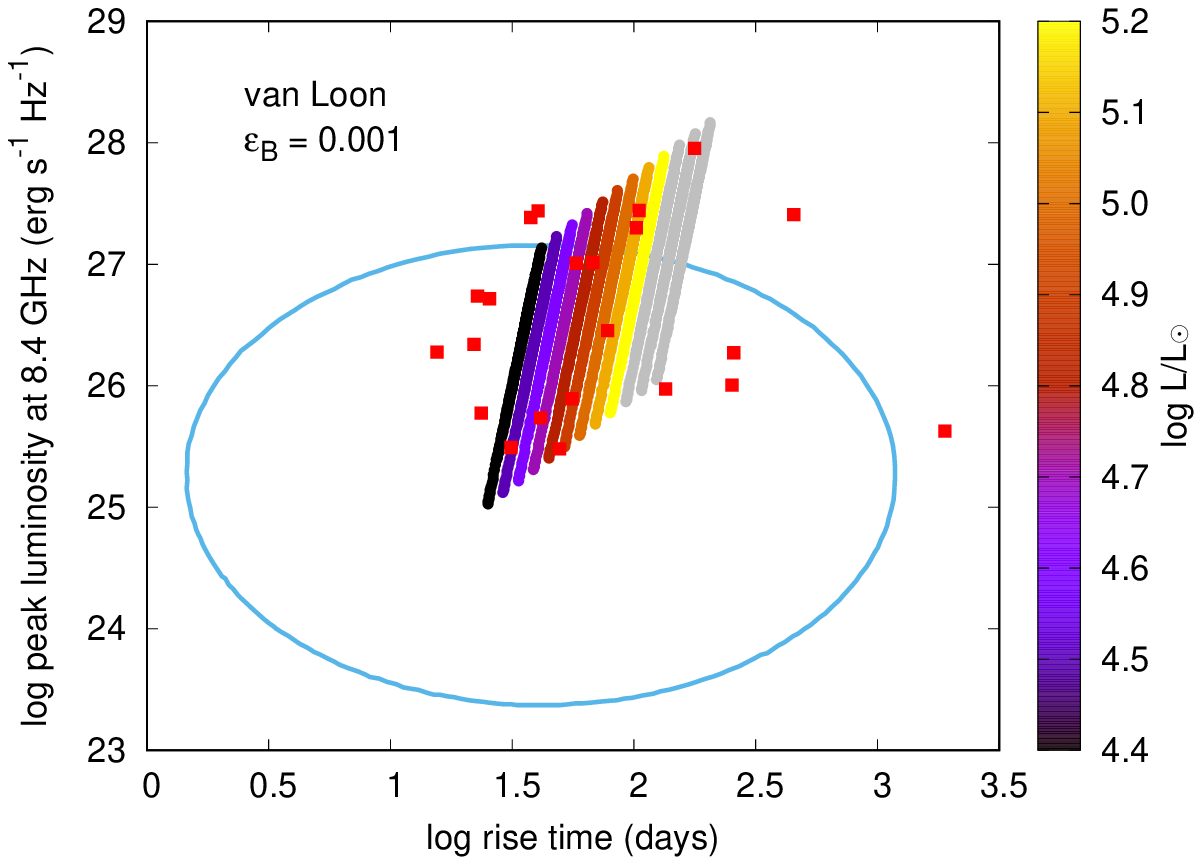}
\caption{
Same as Fig.~\ref{fig:standard}, but for $\varepsilon_B = 0.001$.
}
\label{fig:epsb}
\end{figure}

\section{RSG mass-loss prescriptions}\label{sec:rsgmasslosspre}
Many empirical prescriptions for the RSG mass-loss rate have been proposed \citep[e.g.,][for a summary]{mauron2011}. They can be broadly classified to two types; those with relatively high mass-loss rates as represented by \citet{vanloon2005} and those with relatively low mass-loss rates as represented by \citet{dejager1988}. We show these RSG mass-loss rates in Fig.~\ref{fig:masslossrate}. The empirical mass-loss rates depend on the bolometric luminosity ($L$) and effective temperature of RSGs. Because we investigate the RSG mass loss through SNe, the ranges of the luminosity and effective temperature we can constrain here are those of RSG SN progenitors. The effective temperature of the RSG SN progenitors is around 3500~K \citep[][]{smartt2015} and we assume this temperature for all the RSG SN progenitors. The lowest luminosity of the RSG SN progenitors is observationally constrained to be $\log (L/\Lsun) \simeq 4.4$ \citep[][]{davies2020}, which is the lowest luminosity we adopt in this paper. The maximum luminosity of RSGs is $\log (L/\Lsun) \simeq 5.5$ \citep[][]{davies2018,davies2020}, while the estimated maximum luminosity of the RSG SN progenitors based on the current RSG SN progenitor identification is $\log (L/\Lsun) \simeq 5.2$ \citep[e.g.,][]{smartt2009}. We adopt the RSG luminosity of up to $\log (L/\Lsun) \simeq 5.5$.

\begin{figure}
\includegraphics[width=\columnwidth]{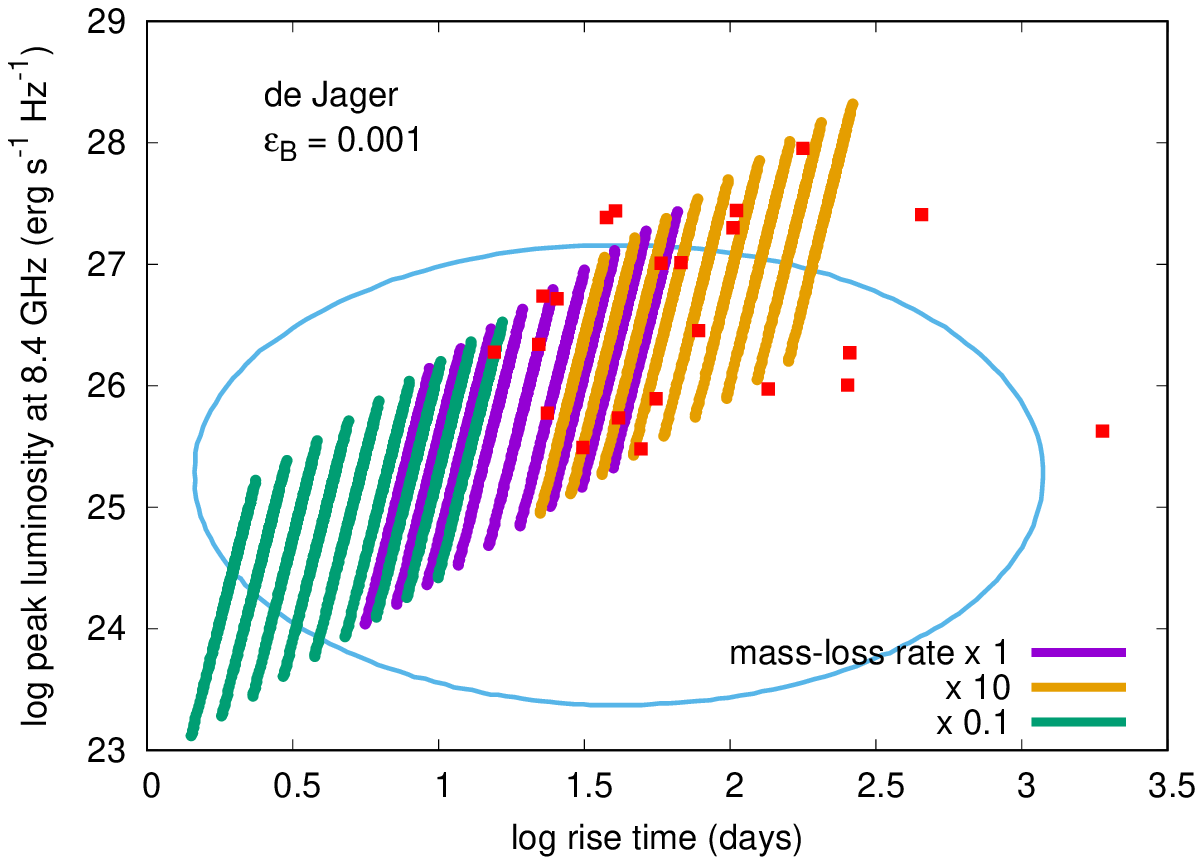}
\includegraphics[width=\columnwidth]{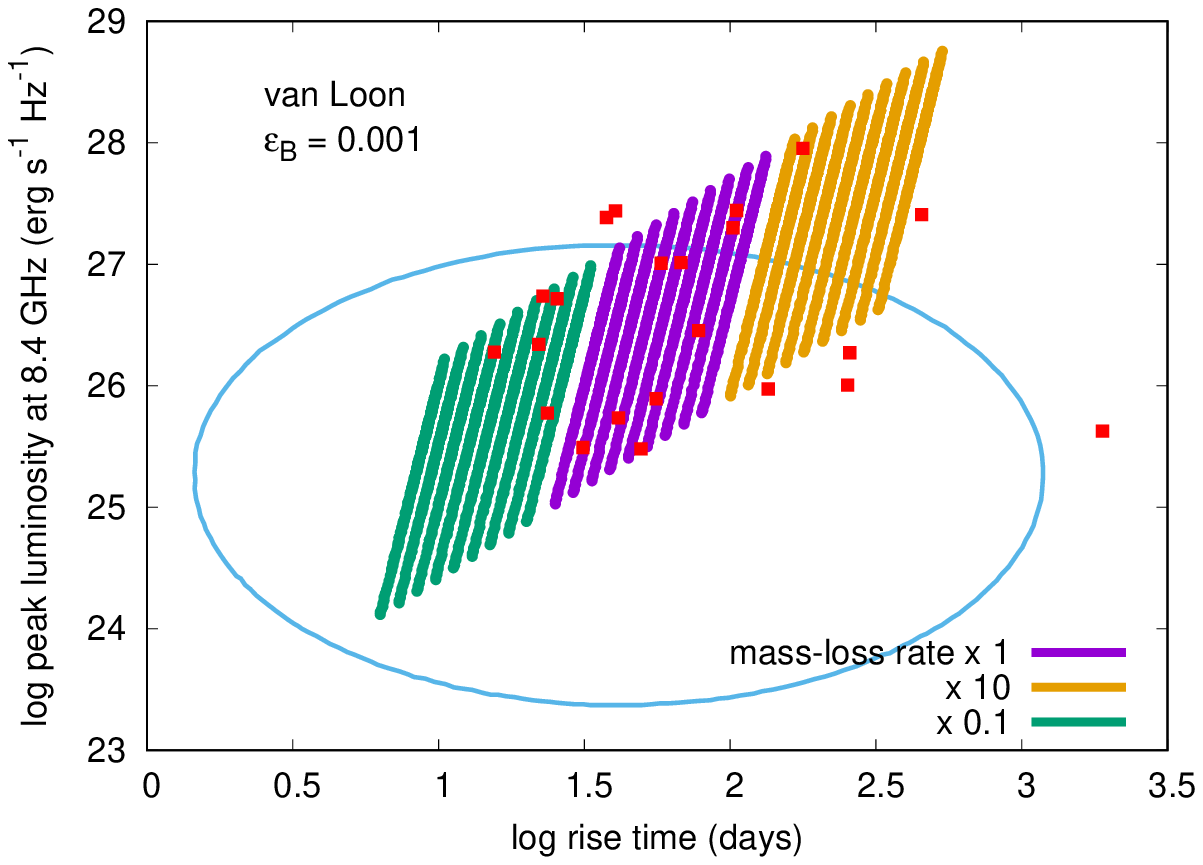}
\caption{
Rise times and peak luminosities estimated by changing the mass-loss rates by a factor of 10 and 0.1. We exclude RSGs with $\log L/\Lsun = 5.3, 5.4,$ and 5.5 in this figure. The top panel is for the de Jager mass-loss rate and the bottom panel is for the van Loon mass-loss rate. $\varepsilon_B = 0.001$ and SSA are assumed.
}
\label{fig:mdot}
\end{figure}

\begin{figure}
\includegraphics[width=\columnwidth]{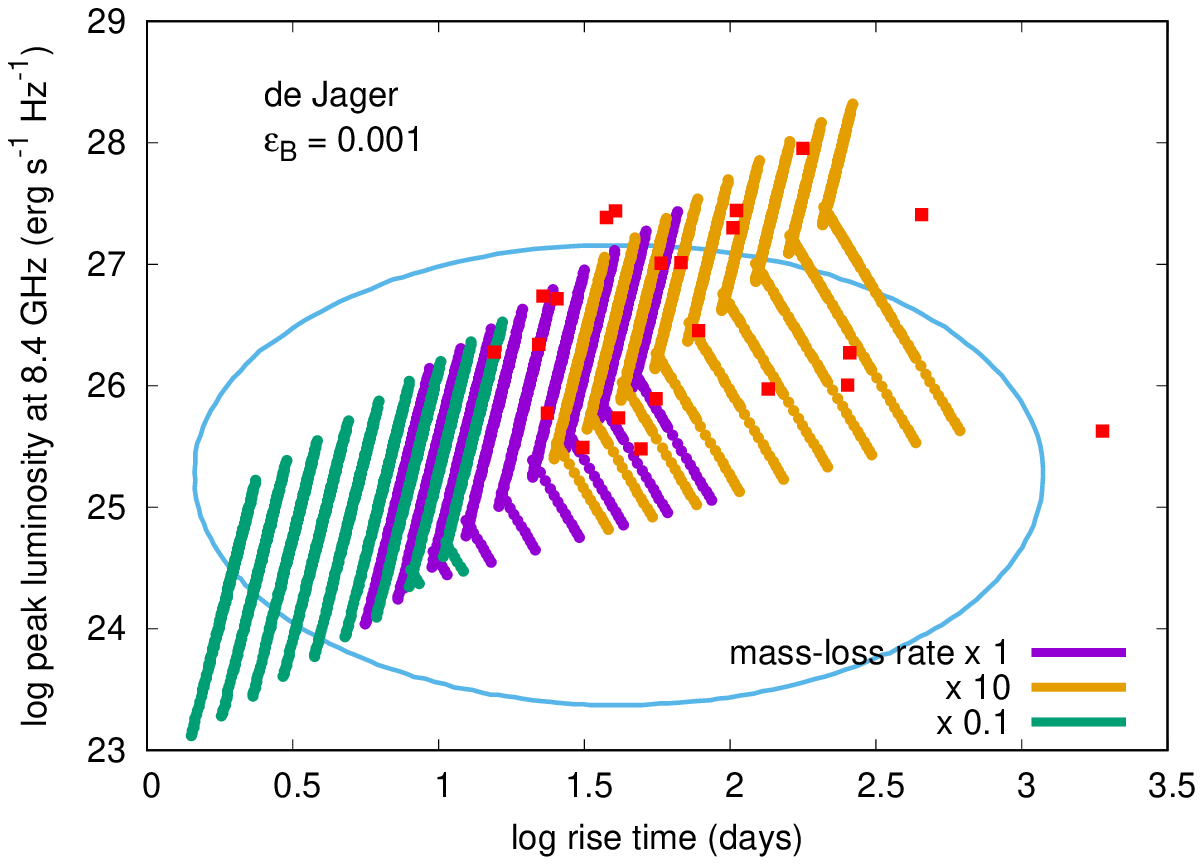}
\includegraphics[width=\columnwidth]{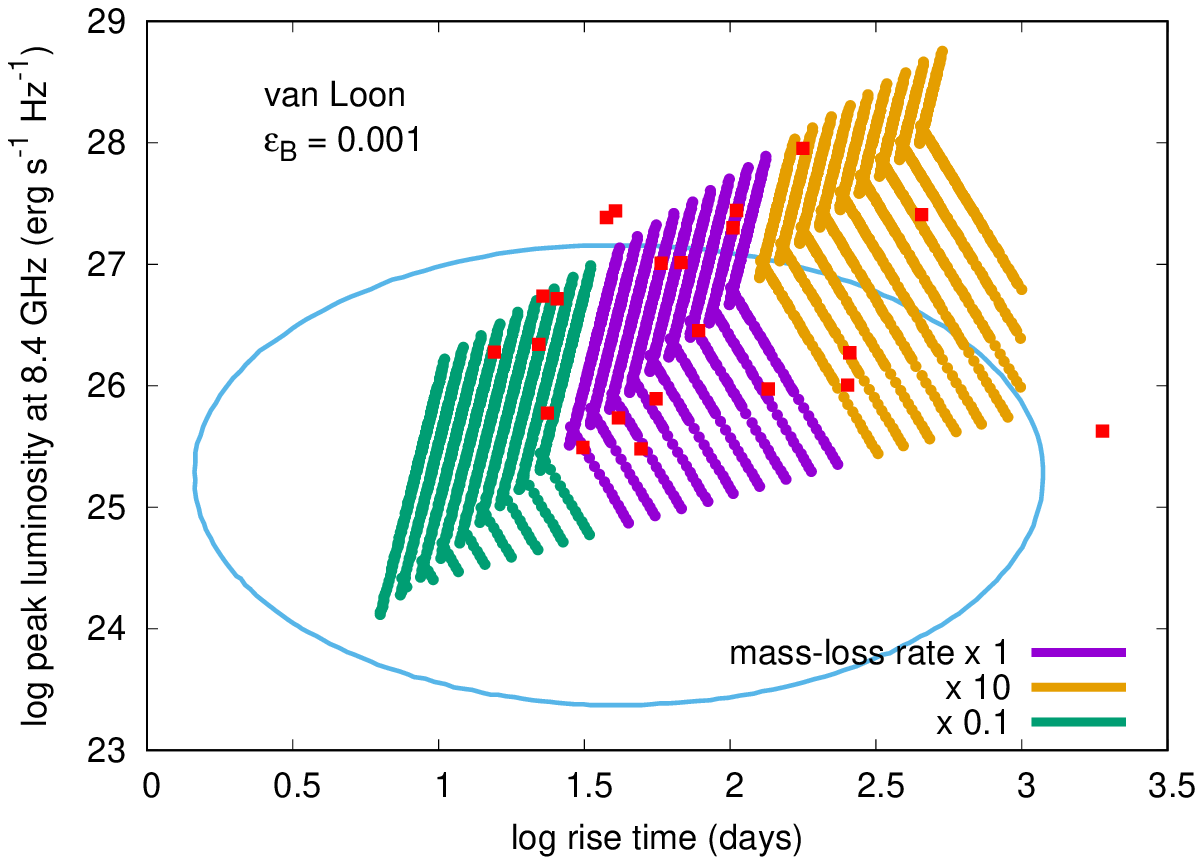}
\caption{
Same as Fig.~\ref{fig:mdot} but the free-free absorption is taken into account.
}
\label{fig:mdot_sf}
\end{figure}

\section{Constraining RSG mass-loss prescriptions with radio SN observations}\label{sec:comparison}
In this section, we estimate the rise time and peak luminosity of SNe~II in radio expected from the RSG mass-loss prescriptions introduced in the previous section and compare the results to the observed distribution.

The top panel of Fig.~\ref{fig:standard} shows the comparison using the de Jager RSG mass-loss rate. Each line shows the rise time and peak luminosity range for a given RSG luminosity with $\Mej = 5-20~\Msun$ and $\Eej = 0.1-2~\mathrm{B}$ (cf. Fig.~\ref{fig:m6example}). The lines for the less luminous RSGs, which have smaller mass-loss rates, locate more left in the rise time and peak luminosity plane. The three gray lines in Fig.~\ref{fig:standard} are for $\log (L/\Lsun)=5.3$, 5.4, and 5.5. As presented before, it is not clear if RSGs with these luminosities explode as SNe. The red squares show the rise times and peak luminosities of SNe~II observed so far \citep[][]{bietenholz2020}. The blue ellipse shows the region where 68\% of SNe~II are estimated to locate in the rise time - peak luminosity plane based on all the observational information including upper limits \citep[][]{bietenholz2020}. Most of the observational points (red squares) locate at the regions that are expected from the de Jager rate, but the theoretical prediction does not reach the low peak luminosity regions included in the blue ellipse. 

The bottom panel in Fig.~\ref{fig:standard} shows the same comparison as the top panel but for the van Loon RSG mass-loss rate. The van Loon mass-loss rate is less dependent on the luminosity than the de Jager mass-loss rate (Fig.~\ref{fig:masslossrate}). Thus, the expected area to cover in the rise time and peak luminosity plane for the van Loon mass-loss rate is smaller than that covered by the de Jagear mass-loss rate. We can also find that the expected rise time and peak luminosity from the van Loon mass-loss rate only cover the largest rise time and peak luminosity ranges found from the observations. Overall, the van Loon mass-loss rate does not explain the widely spread rise time and peak luminosity relation in SNe~II, and it does not explain the low peak luminosity that dominates SNe~II.

In both de Jager and van Loon mass-loss rates, we found that the theoretical predictions do not reach the low peak luminosity expected from the observations shown as the blue ellipse in Fig.~\ref{fig:standard}. This could be partly because of our assumption of $\varepsilon_B = 0.01$. As discussed before, $\varepsilon_B$ is uncertain and a smaller $\varepsilon_B$ can make the expected peak luminosity lower. Fig.~\ref{fig:epsb} shows the same estimates as in Fig.~\ref{fig:standard} but for $\varepsilon_B=0.001$. We can find that the rise time and peak luminosity expected from the de Jager rate starts to cover some low peak luminosity region, while the van Loon mass-loss rate still does not explain the low peak luminosities.

In addition to the uncertainties in physics, we need to take the diversity in the RSG mass-loss rates into account. Although the mass-loss prescriptions provide one mass-loss rate for one luminosity, dispersion within a factor of around 10 in the RSG mass-loss rate is known to exist \citep[e.g.,][]{mauron2011,goldman2017}. Fig.~\ref{fig:mdot} shows the expected rise time and peak luminosity regions with the mass-loss rate within a factor of 10 from the de Jager and van Loon prescriptions in the case of $\varepsilon_B=0.001$. The de Jager mass-loss rate reproduces the diverse rise time and peak luminosity estimated from the observations, while the van Loon mass-loss rate predict larger rise time and higher luminosity than those observed in SNe~II. The region covered in the rise time and peak luminosity plane is not as widely spread as that covered by the de Jager mass-loss rate. As discussed before, this is because the van Loon mass-loss rate does not strongly depend on luminosity and the CSM density around RSG SN progenitors is not expected to have much diversity.

We have not taken the effect of the free-free absorption into account so far. The optical depth to the free-free absorption at 8.4~GHz becomes unity at \citep[][]{chevailer2006}
\begin{equation}
    t_\mathrm{ff} \simeq 6\left(\frac{\dot{M}/10^{-6}~\Msunpyr}{v_\mathrm{CSM}/10~\kmps} \right)^{2/3}\left(\frac{T_\mathrm{CSM}}{10^5~\mathrm{K}}\right)^{-1/2}\left(\frac{v_\mathrm{sh}}{10^4~\kmps}\right)^{-1}~\mathrm{days},
\end{equation}
where $T_\mathrm{CSM}$ is the CSM temperature. $T_\mathrm{CSM}$ is uncertain, but we adopt $T_\mathrm{CSM}\simeq 10^5~\mathrm{K}$ in this work \citep[][]{lundqvist1988,chevailer2006}. If the time when $\tau_\mathrm{SSA}$ becomes unity is smaller than $t_\mathrm{ff}$, the free-free absorption determines the radio LC peak. Fig.~\ref{fig:mdot_sf} shows the rise time - peak luminosity relations where the free-free absorption is taken into account. We can find that the models with large $\dot{M}$ are strongly affected by free-free absorption. They show a break in the rise time - peak luminosity relation and the region with the long rise time and low peak luminosity (the right bottom part of the rise time - peak luminosity plane) that is not covered by the SSA models is covered by taking the free-free absorption into account. Our conclusion that the de Jager rate explains the widely spread rise time and peak luminosity relation better, however, remains unchanged.

\section{Summary}\label{sec:summary}
We have compared the rise time and peak luminosity distribution of SNe~II in radio with those expected from the RSG mass-loss prescriptions of \citet{dejager1988} and \citet{vanloon2005}. We found that the de Jager mass-loss rate explains the rise time and peak luminosity of SNe~II in radio well considering the dispersion in the RSG mass-loss rates (Fig.~\ref{fig:mdot_sf}). The van Loon mass-loss rate generally predicts longer rise times and higher peak luminosities than those found in SNe~II. This is because of the relatively high mass-loss rates that the van Loon prescription provides. The uncertainty in physics and the dispersion in the RSG mass-loss rates are not likely to fully resolve this discrepancy.

Another superiority of the de Jager mass-loss rate in explaining the radio properties of SNe~II is that it can explain the wide range of the rise times and peak luminosities found in SNe~II. The de Jager mass-loss rate can reproduce the wide range because of its steep dependence on luminosity that allows it to produce a variety of mass-loss rates with the RSG SN progenitor luminosity range (Fig.~\ref{fig:masslossrate}). The de Jager mass-loss rate changes by a factor of around 100 from the lowest luminosity RSG SN progenitors to the highest, while the van Loon mass-loss rate changes only by a factor of around 10. The RSG mass-loss prescriptions having as strong luminosity dependence as the de Jager prescription is preferred to explain the diversity in SN~II radio properties.

RSG mass-loss rates are an essential ingredient to understand the evolution of massive stars towards SN explosions and SN properties. We have shown in this paper that radio SN observations can provide a way to constrain the uncertain RSG mass-loss prescriptions. The radio SN observations are still limited and many more observations are required to constrain the rise time and peak luminosity of SNe in radio. Further radio observations would lead to a better understanding of mass loss in massive stars.


\section*{Acknowledgements}
I would like to thank the anonymous referee for the helpful comments that improved this work.
This work is supported by the Grants-in-Aid for Scientific Research of the Japan Society for the Promotion of Science (JP18K13585, JP20H00174).

\section*{Data Availability}
The data underlying this article will be shared on reasonable request to the corresponding author.



\bibliographystyle{mnras}
\bibliography{references} 






\bsp	
\label{lastpage}
\end{document}